\documentclass[aps,prd,10pt,amsmath,floats,floatfix,twocolumn,nofootinbib,superscriptaddress,preprintnumbers 
]{revtex4-1}

\usepackage{graphicx}
\usepackage{amssymb}
\usepackage{color}
\usepackage{enumerate}
\usepackage[breaklinks]{hyperref}
\usepackage[utf8]{inputenc}
\usepackage[normalem]{ulem}

\usepackage{latexsym}
\usepackage{amssymb,amsthm}
\usepackage{bm}

\newtheorem{thm}{Theorem}
\newtheorem{cor}{Corollary}

\newcommand{\beaa}{\begin{equation} \begin{array}{ll}}
\newcommand{\eeaa}{\end{equation} 	\end{array} }
\def\bs{\begin{subequations}}
\def\es{\end{subequations}}
\newcommand{\Eq}[1]{(\ref{#1})}

\def\cob{\color{blue}}
\newcommand{\book}[5]{\emph{#1} (#2, #3, #4, #5)}

\newcommand{\oarX}[1]{\href{http://arxiv.org/abs/#1}{{\ttfamily\cob arXiv:#1}}}
\newcommand{\arX}[1]{\href{http://arxiv.org/abs/#1}{{\ttfamily\cob arXiv:#1}}}
\newcommand{\doin}[6]{\href{http://dx.doi.org/#1}{{\cob #2 #3 {\bf #4}, #5 (#6)}}}
\newcommand{\doinn}[5]{\href{http://dx.doi.org/#1}{{\cob #2 {\bf #3}, #4 (#5)}}}
\newcommand{\doij}[5]{\href{http://dx.doi.org/#1}{{\cob #2 #3 (#5) #4}}}

\newcommand{\ndoinn}[5]{\href{#1}{{\cob #2 {\bf #3}, #4 (#5)}}}
\newcommand{\procsinm}[5]{in \emph{#1}, ed.\ by #2 (#3, #4, #5)}
\newcommand{\tia}[1]{#1,}

\newcommand{\be}{\begin{eqnarray}}
\newcommand{\ee}{\end{eqnarray}}
\newcommand{\ba}{\begin{eqnarray}}
\newcommand{\ea}{\end{eqnarray}}

\usepackage{tikz}
\newcommand*\circled[1]{\tikz[baseline=(char.base)]{
            \node[shape=circle,draw,inner sep=1pt] (char) {#1};}}

\def\a{\alpha}
\def\b{\beta}

\def\g{\gamma}

\def\De{\Delta}

\def\s{\sigma}

\def\p{\partial}

\def\B{\Box}

\def\lp{\ell_{\rm Pl}}

\def\rme{e}
\def\rmd{d}

\begin{document}

\title{Nonlinear stability in nonlocal gravity} 

\author{Fabio Briscese}
\affiliation{SUSTech Academy for Advanced Interdisciplinary Studies, Southern University of Science and Technology, Shenzhen 518055, China}
\affiliation{Istituto Nazionale di Alta Matematica Francesco Severi, Gruppo Nazionale di Fisica Matematica, Citt\`{a}
	Universitaria, P.le A.\ Moro 5, 00185 Rome, Italy}

\author{Gianluca Calcagni}
\affiliation{Instituto de Estructura de la Materia, CSIC, Serrano 121, 28006 Madrid, Spain}

\author{Leonardo Modesto}
\affiliation{Department of Physics, Southern University of Science and Technology, Shenzhen 518055, China} 

\date{January 11, 2018}

\begin{abstract}
We address the stability issue of Ricci-flat and maximally symmetric spacetimes in nonlocal gravity to all perturbative orders in the gravitational perturbation. Assuming a potential at least cubic in curvature tensors but quadratic in the Ricci tensor, our proof consists on a mapping of the stability analysis in nonlocal gravity to the same problem in Einstein--Hilbert theory. One of the consequences is that only the graviton field can propagate and the theory is ghost-free at all perturbative orders. All the results known in Einstein gravity in vacuum with or without a cosmological constant can be exported to the case of nonlocal gravity: if a spacetime is stable at all perturbative orders in Einstein gravity, it is stable also in nonlocal gravity. Minkowski and de Sitter spacetimes are particular examples. We also study how the theory affects the propagation of gravitational waves in a cosmological background.
\end{abstract}

\preprint{\doin{10.1103/PhysRevD.99.084041}{PHYSICAL REVIEW}{D}{99}{084041}{2019} \hspace{8.5cm} \arX{1901.03267}}

\maketitle


\section{Introduction}

The golden age of gravitational-wave (GW) observations, from those emitted by small-redshift astrophysical compact objects \cite{Ab16a,Ab17b} to the inflationary tensor spectrum generated in the early universe \cite{P18I}, has consecrated some among the most notable achievements of Einstein's general relativity. To date, its predictions on those small ripples of spacetime known as GWs have been confirmed and no evidence for new physics has been found. While with the available LIGO data we are closing in on many models beyond general relativity, new astrophysical and cosmological constraints are devised to be tested in near-future experiments such as Euclid and LISA, just to mention two.

Among the things that could happen ``beyond general relativity,'' the classical dynamics of spacetime and matter may be modified, or gravity may go quantum and the laws of quantum mechanics affect the texture of spacetime. Among the many candidates for quantum gravity, nonlocal quantum gravity (NLG) has been developing at a fast pace \cite{modesto,BCKM,CaMo2,modestoLeslaw,universality,Review,Buoninfante:2018lnh}. Motivations about this theory and a description of its main features were already given in past literature, but we recapitulate them here for the unfamiliar reader. ``Quantum gravity'' is a family of proposals expected not only to unify quantum mechanics and the gravitational force in a consistent theory, but also to address some pending issues of general relativity, in particular, the presence of singularities. The main reason of interest in the nonlocal quantum theory considered in this paper is that it is renormalizable (or finite, in some of its incarnations) and unitary, two requisites that any candidate of quantum gravity should possess to be theoretically viable. These properties are not found in the quantization of many modified gravity models, where the dynamics of classical Einstein gravity is deformed \emph{ad hoc} for the purpose of generating interesting phenomenology. Quantizing gravity may result in intricate formalisms, but this theory in particular is not especially complicated. Recall that the very well known local Stelle gravity (quadratic gravity \cite{Ste77,Ste78}) is renormalizable but plagued by ghost instabilities, due to the presence of fourth-order derivatives. The fact that introducing an entire nonlocal operator in the same action guarantees unitarity makes the nonlocal approach parsimonious and, to some of us, attractive. 

This field theory, unitary and finite at the quantum perturbative level, stems from early proposals by Krasnikov \cite{Krasnikov} and Kuz'min \cite{kuzmin} (see also \cite{Tom97,Sie03,Kho06}). In particular, there is a finite completion of the theory which does not show ultraviolet divergences at any order in the loop expansion \cite{modestoLeslaw,universality}, while perturbative unitarity based on the Cutkosky cutting rules has been proved at any perturbative order in the loop expansion \cite{PiSe,brisceseUnitarity,ChTo}. Another promising theory that deserves to be mentioned is Lee--Wick quantum gravity \cite{shapiromodesto,LWqg}, a special case of \cite{shapiro3}. This theory has complex conjugate poles beside the graviton field, but such poles can be consistently removed from the physical spectrum and never go on shell. 

For the nonlocal models considered here, the classical solutions of Einstein gravity in vacuum are also solutions of NLG \cite{yaudong} and the linear stability analysis in NLG is the same as in Einstein gravity \cite{CM,CMM}. Indeed, Ricci-flat spacetimes in NLG are stable under linear perturbations if they are stable in Einstein gravity \cite{CMM}. Just as in general relativity \cite{WR}, Schwarzschild spacetime is a stable solution in NLG at the linear level. Whether this solution also represents an actual, astrophysical black hole remains to be seen, since approximated solutions of the linearized equations of motion (EOM) look regular \cite{Tse95,FZdP,Fro15,Fro16,Gia1,Gia2}. More recently, in \cite{BrMo} it was found that the NLG dynamics of small perturbations of Minkowski metric is the same as in Einstein gravity. The stability of Minkowski spacetime in NLG was thus inferred from its stability in general relativity at the nonlinear level, i.e., to all perturbative orders in the gravitational perturbation (see \cite{ChKl,Chr91,LiRo,Lin17,Briscese:2017vff} for the stability of small perturbations of the Minkowski metric in general relativity).

In this paper, we will address the stability issue in NLG for a larger class of spacetimes, including all Ricci-flat and maximally symmetric ones. Following the same line of reasoning outlined in \cite{BrMo}, we will show that small perturbations of Ricci-flat and maximally symmetric metrics in NLG satisfy the same equations of motion as in Einstein theory. The conclusion, which will have a deep impact on the original question about the phenomenological applications of quantum gravity, is that Ricci-flat and maximally symmetric spacetimes are stable in NLG \emph{to all perturbative orders} if they are stable in Einstein gravity. Also to all orders, no new degrees of freedom propagate. These results include backgrounds important for cosmology such as (anti--)de Sitter, for which stability and the absence of extra propagating degrees of freedom was known at the linear order \cite{BKM,KSML}.

In  Sec.\ \ref{Nonlocal Gravity without cosmological constant}, we will prove our result for Ricci-flat spacetimes, while the case of maximally symmetric metrics will be addressed in Sec.\ \ref{Nonlocal Gravity with cosmological constant}. Section \ref{concl} summarizes the main results and outlines their present and future physical applications. For an easier reading, most of the calculations have been confined into Appendices \ref{app1} and \ref{app2}.

Our conventions are that the metric tensor $g_{\mu \nu}$ has signature $(- + \dots +)$ and the curvature tensors are defined as $R^{\mu}_{\ \nu \rho \sigma} = - \partial_{\sigma} \Gamma^{\mu}_{\nu \rho} + \dots $, $R_{\mu \nu} = R^{\rho}_{\ \mu  \rho \nu}$ and $R = g^{\mu \nu} R_{\mu \nu}$. Terms quadratic in the Ricci tensor or scalar but not in the Riemann tensor will be denoted as $O({\bf Ric}^2)$.

\section{Nonlocal gravity without cosmological constant} \label{Nonlocal Gravity without cosmological constant}

A general class of theories compatible with unitarity and superrenormalizability or finiteness is, in $D$ dimensions \cite{CMM},
\ba
S &=& \frac{1}{2\kappa^2}\!\!\int\!\! \rmd^D x \sqrt{|g|}[R + R \gamma_0 (\Delta_{\rm L}) R + R_{\mu\nu} \gamma_2(\Delta_{\rm L}) R^{\mu\nu}\nonumber\\
&& \qquad\qquad+ V_g],\label{action}
\ea
where the ``potential'' term $V_g$  is built with the curvature tensor and it is at least quadratic in the Ricci tensor, as required for superrenormalizability or finiteness. However, in what follows we will consider potentials at least cubic in curvature tensors but quadratic in the Ricci tensor, e.g., $V_g= R_{\alpha \beta} R^{\beta \gamma} R^\alpha_{\,\,\,\gamma}+R^{\alpha \rho} R^{\beta \gamma} R_{\a\b\rho\gamma}+O({\bf Ric}^n\times {\bf Riem}^m)$, with $n\geq 2$ and $n+m\geq 3$, and we will show that under this restriction we can prove the stability of Ricci-flat solutions. The reader interested in the form and role of cubic potentials can consult Ref.~\cite{modestoLeslaw}. The $\gamma_{0,2}(\Delta_{\rm L})$ in Eq.~(\ref{action}) are the nonlocal form factors, which are functions of the Lichnerowicz operator $\Delta_{\rm L}$. When acting on a rank-2 symmetric tensor, $\Delta_{\rm L}$ is defined as
\be
\Delta_{\rm L} X_{\mu\nu} = 2 R^\s\,_{\mu\nu\tau} \, X^\tau\,_\s + R_{\mu \s} \, X^\s\,_\nu + R_{\s\nu} \, X^\s\,_\mu  - \Box X_{\mu\nu} \nonumber
\ee
On the trace $X^\mu_\mu$ or on a scalar $X$, $\Delta_{\rm L}X=-\Box X$.
 
The EOM in vacuum for the action (\ref{action}) in a compact and short notation read \cite{Mirzabekian:1995ck} 
\ba
0&=&
 G_{\mu\nu} - \frac{1}{2} g_{\mu\nu} R \gamma_0(\Delta_{\rm L}) R -  \frac{1}{2} g_{\mu\nu} R_{\alpha \beta} \gamma_2(\Delta_{\rm L}) R^{\alpha \beta}
\nonumber\\&&
+2\frac{\delta R}{\delta g^{\mu \nu}} \gamma_0(\Delta_{\rm L})R+ \frac{\delta R_{\alpha \beta}}{\delta g^{\mu \nu}} \gamma_2(\Delta_{\rm L})R^{\alpha \beta}  \nonumber \\
&& 
+ \frac{\delta R^{\alpha \beta}}{\delta g^{\mu \nu}  } \gamma_2(\Delta_{\rm L})R_{\alpha \beta} +  \frac{\delta \Delta_{\rm L}^r}{\delta g^{\mu\nu} }
 \left[\frac{\gamma_0(\Delta_{\rm L}^l)-\gamma_0(\Delta_{\rm L}^r)}{\Delta_{\rm L}^r - \Delta_{\rm L}^l} R R \right]
 \nonumber \\
&& 
 + \frac{\delta \Delta_{\rm L}^r}{\delta g^{\mu\nu}}\left[
  \frac{ \gamma_2(\Delta_{\rm L}^l)- \gamma_2(\Delta_{\rm L}^r)}{\Delta_{\rm L}^r - \Delta_{\rm L}^l} R_{\alpha \beta} R^{\alpha \beta} \right]+ \frac{\delta {V_g}}{ \delta g^{\mu \nu}}\,,\label{EOM}
\ea
where $\Delta_{\rm L}^{l,r}$ act, respectively, on the left and right arguments inside the brackets.

In order to have a ghost-free theory (tree-level unitarity when we expand the action around the Minkowski vacuum) without the Starobinsky curvaton mode \cite{BMMS,BMT}, we are forced to select \cite{Kho06,Modesto:2013ioa}
\be\label{gaga}
\gamma_0= -\frac{\gamma_2}{2}\,,
\ee
where $\g_2=\g$ and
\be\label{LicForm}
\gamma (\Delta_{\rm L}) = \frac{\rme^{{\rm H}(\sigma \Delta_{\rm L})} - 1}{- \Delta_{\rm L}} \,.
\ee
Here ${\rm H}$ is an analytic entire function whose properties are dictated by the superrenormal\-iza\-bil\-ity of the theory \cite{modesto,modestoLeslaw} and $\sigma$ is a parameter that fixes the length scale of nonlocality $l \equiv \sqrt{\sigma}$. Moreover, we note that, since $V_g$ is at least cubic in the Ricci curvature, its variation with respect to the metric tensor will be at least quadratic, so we can write $\frac{\delta {V_g}}{\delta g^{\mu \nu}} \propto {\bf Ric}^2$. Therefore, Eq.\ (\ref{EOM}) can be recast in the following form \cite{CMM}:
\be
G_{\mu\nu}+2 \frac{\delta R_{\alpha \beta}}{\delta g^{\mu \nu}} \gamma (\Delta_{\rm L}) G^{\alpha \beta} + O({\bf Ric}^2)_{\mu\nu} =0\, , 
\label{EOM1}
\ee
where the variation of the Ricci tensor is given in Eq.~\Eq{varric} and $O({\bf Ric}^2)$ stands for operators at least quadratic in the Ricci tensor. 

From the cumbersome computation developed in \cite{CMM}, it turns out that the functional dependence of the form factor (\ref{LicForm}) on the Lichnerowicz operator leads to a simplified form of the EOM (\ref{EOM1}), 
\be
\mathcal{E}_{\mu\nu}:=\rme^{{\rm H}(\sigma \Delta_{\rm L})} G_{\mu\nu} + \mathcal{Q} ({\bf Ric})_{\mu\nu}=0\,,
\label{LastEOM}
\ee
where $\mathcal{Q}_{\mu\nu}$ is the sum of terms at least quadratic in the Ricci tensor, e.g., $\sigma  \left[\left(\sigma\Box\right)^n R_{\mu\alpha}\right]\left[\left(\sigma\Box\right)^m R^{\alpha}_{\,\,\,\nu}\right]$ or $\sigma^2 \left[\left(\sigma\Box\right)^n  R_{\mu\alpha}\right]\left[\left(\sigma\Box\right)^m R^{\alpha}_{\,\,\,\nu}\right][\left( \sigma \Box\right)^l R]$, for integer $n, m, l$.

It is straightforward to show that Eq.\ (\ref{LastEOM}) is satisfied if $R_{\mu\nu} =0$,  so that all Ricci-flat spacetimes are solutions of the EOM. In particular, the Schwarzschild metric, the Kerr metric, and all the known Ricci-flat metrics in Einstein gravity without matter are exact solutions of the nonlocal theory too.

Finally, we can move the exponential operator in Eq.\ (\ref{LastEOM}), from the left-hand to the right-hand side of the EOM and replace the Ricci tensor with the Einstein tensor, so that we find the following structure for the final, but implicit, EOM (\ref{EOM}):
\be
\hspace{-0.5cm}
{\bf G} =  \rme^{- {\rm H}(\sigma \Delta_{\rm L})} \, \mathcal{Q} (
{\bf G} ) = \rme^{- {\rm H}(\sigma \Delta_{\rm L})} \, \left(
{\bf G} \, \mathcal{Q}_2 \,  {\bf G} +\ldots \right), 
\label{LastEOMG}
\ee
where ${\bf G}$ is the Einstein tensor $G_{\mu\nu}$ and the term ${\bf G} \, \mathcal{Q}_2 \,  {\bf G} $ is quadratic in $G_{\mu\nu}$. In our notation, $\mathcal{Q}_2$ is the sum of operators acting on the left and right ${\bf G}$, while the indices in the two ${\bf G}$ are contracted to form a rank-2 covariant tensor. The ellipsis in Eq.\ (\ref{LastEOMG}) includes all the terms in $\mathcal{Q}( {\bf G})$ at least cubic in ${\bf G}$, which can be neglected in our discussion of the stability, as explained in detail below.  All the operators $\mathcal{Q}({\bf G})$ can be explicitly and easily derived from the EOM (\ref{EOM}). However, in order to address the stability problem we only need to know that such operators are at least quadratic in the Einstein tensor. 

Now we are ready to present the following theorem. 
\begin{thm}
In nonlocal gravity (\ref{action})-\Eq{gaga}-\Eq{LicForm} and in vacuum {\rm (}${\bf T}=0$, where ${\bf T}$ is the matter energy-momentum tensor{\rm )}, the gravitational perturbations of Ricci-flat spacetimes satisfy the same EOM of the perturbations in Einstein gravity in vacuum.
\end{thm}
Two different proofs are given in Appendix \ref{app1}. In the first, we consider a Ricci-flat metric $g^{(0)}_{\mu\nu}$, that is, a metric such that $R_{\mu\nu}(g^{(0)}) = 0$. As discussed before, $g^{(0)}_{\mu\nu}$ is also a solution of the EOM (\ref{LastEOMG}) for nonlocal gravity. Small perturbations around the background $g^{(0)}_{\mu\nu}$ are written as an expansion of the full metric tensor $g_{\mu\nu}$ in powers of a small parameter $\epsilon\ll 1$ as  
\be
g_{\mu\nu} = \sum_{n=0}^{\infty} \epsilon^n h^{(n)}_{\mu\nu} \, ,  \qquad h^{(0)}_{\mu\nu} \equiv g^{(0)}_{\mu\nu} \, .
\label{hEXP}
\ee
Then, it is showed (see Proof 1 in Appendix \ref{app1}) that at any perturbative order $n>0$ in $\epsilon$ 
\be
{\bf \mathcal{E}}^{(n)} = 0 \quad \Longrightarrow \quad {\bf G}^{(n)}=0 \, ,
\ee
where ${\bf \mathcal{E}}^{(n)}$ (${\bf G}^{(n)}$) are the nonlocal (respectively, local) EOM (\ref{LastEOM}) (respectively, Einstein EOM) at the order $\epsilon^n$, so that
\be
&& \hspace{-.6cm}{\mathcal{E}}_{\mu\nu}(g_{\mu\nu}) = \sum_{n=0}^\infty \epsilon^n {\mathcal{E}}_{\mu\nu}^{(n)}=0,\quad {\mathcal{E}}_{\mu\nu}^{(0)} \equiv {\mathcal{E}}_{\mu\nu}(g^{(0)}) = 0 \, ,
\nonumber \\
%
&& \hspace{-.6cm}G_{\mu\nu}(g_{\mu\nu}) = \sum_{n=0}^\infty \epsilon^n G_{\mu\nu}^{(n)}=0, \,\,\, G_{\mu\nu}^{(0)} \equiv G_{\mu\nu}(g^{(0)}) = 0 \, .
\label{EoM Perturbations}
\ee
In other words, the dynamics of small perturbations in NLG is the same as in Einstein gravity. The second proof (see Proof 2 in Appendix \ref{app1}) starts from a different but equivalent Lagrangian that involves an auxiliary rank-2 traceless tensorial field $\phi_{\mu\nu}$ and a scalar field $\chi$. We are interested in this other demonstration because it clearly shows that these new degrees of freedom encoded in NLG do not propagate at any perturbative order in vacuum. This powerful conclusion extends, both to more general backgrounds and to all perturbative orders, the well-known result that the only propagating degree of freedom of this theory in Minkowski is the graviton. It also explains why the theory is exactly unitary, contrary to Stelle gravity where the spin-2 ghost field propagates already at the first-order perturbative level \cite{Ste77,Ste78}.

The stability of Ricci-flat spacetimes follows from our theorem:
\begin{cor}
If a Ricci-flat solution is stable in Einstein gravity, it is also stable in NLG (\ref{action})-\Eq{gaga}-\Eq{LicForm}.
\end{cor}
This result stems directly from the fact that the evolution of small perturbations of Ricci-flat solutions is the same in the two theories. For instance, Minkowski spacetime is stable, since it is stable in general relativity \cite{BrMo}, and the same will be true for other Ricci-flat spacetimes, e.g., for Schwarzschild black holes, provided they are stable in Einstein gravity.

\section{Nonlocal gravity with cosmological constant} \label{Nonlocal Gravity with cosmological constant}

We now introduce the cosmological constant $\Lambda$ in the theory (\ref{action}) in such a way that maximally symmetric solutions of Einstein's gravity are still solutions in NLG. The action reads
\ba
S &=& \frac{1}{2\kappa^2}\!\!\int\!\! \rmd^D x \sqrt{|g|} \left[
R-2\Lambda 
+ (R - 4 \Lambda ) \gamma_0 (\Delta_\Lambda) ( R - 4 \Lambda )
\right.  \nonumber\\&&
\left.
+ (R_{\alpha \beta}- \Lambda \, g_{\alpha \beta})   \gamma_2(\Delta_\Lambda)  (R^{\alpha \beta}  - \Lambda \, g^{\alpha \beta}) + V_{E}\right],\label{actionLambda}
\ea
where $\Lambda$ is the cosmological constant and $V_{E}$ is a potential at least cubic in the tensor
\be
E_{\mu\nu} := G_{\mu\nu} + \Lambda \, g_{\mu\nu}. 
\label{EinsteinLambda}
\ee
One can equivalently rewrite the theory (\ref{actionLambda}) in terms $E_{\mu\nu}$. 
Finally, the operator $\Delta_\Lambda$ is a generalization of the Lichnerowicz operator $\Delta_{\rm L}$. When acting on a rank-2 symmetric tensor,
\ba
\Delta_\Lambda X_{\mu\nu} &=& 2 R^\s\,_{\mu\nu\tau} \, X^\tau\,_\s 
+ (R_{\mu \s} - \Lambda \, g_{\mu \sigma}) \, X^\s\,_\nu  \nonumber \\
&&+ (R_{\s\nu} - \Lambda \, g_{\sigma \nu}) \, X^\s\,_\mu  - \Box X_{\mu\nu}\,.
 \label{lichLambda}
\ea
The EOM resulting from the variation of the action (\ref{actionLambda}) up to quadratic orders in $E_{\mu\nu}$, which we rename ${\bf E}$, and assuming again the relation \Eq{gaga}, are
\ba
E_{\mu\nu}^{\rm NL} &:=& E_{\mu\nu} + 2 \frac{\delta (R_{\alpha \beta} -  \Lambda \, g_{\alpha \beta}) }{\delta g_{\mu\nu}}    \gamma_2(\Delta_\Lambda)  E^{\alpha \beta} + O({\bf E}^2)\nonumber\\
& =&0 \, .\label{EOML}
\ea
Clearly, all maximally symmetric spacetimes are exact solutions of the theory:
\be
E_{\mu \nu} = 0 \quad \Longrightarrow \quad E_{\mu\nu}^{\rm NL} =0 \, .
\ee
In Appendix \ref{app2}, we show that assuming Eq.~\Eq{gaga} the EOM \Eq{EOML} can be written explicitly as
\be
\Big[1 -(\Delta_\Lambda+4\Lambda)\, \gamma(\Delta_\Lambda) \Big] E_{\mu\nu} + O({\bf E^2}) = 0\,.
\label{EOMLambda5}
\ee
To secure the stability of the theory we define an ${\rm H}$ such that $\g$ is entire:
\be\label{gadel}
\gamma(\Delta_\Lambda) = \frac{\rme^{{\rm H}[\sigma (\Delta_\Lambda + 4 \Lambda)]} - 1}{- (\Delta_\Lambda + 4 \Lambda)},
\ee
and the EOM (\ref{EOMLambda5}) simplify to
\ba
{\bf E} &=&  \rme^{- {\rm H}[\sigma (\Delta_\Lambda + 4 \Lambda)]} \, \mathcal{Q} (
{\bf E} )\nonumber\\
&=& \rme^{- {\rm H}[\sigma (\Delta_\Lambda + 4 \Lambda)]} \, \left({\bf E} \, \mathcal{Q}_2 \,  {\bf E} +\cdots \right)=0  \,.
\label{EOMLF}
\ea
We see that Eq.\ (\ref{EOMLF}) is the same as (\ref{LastEOMG}) with the replacements ${\bf G} \rightarrow {\bf E}$ and $\Delta_{\rm L }\rightarrow \Delta_\Lambda + 4 \Lambda$. Therefore, we can enunciate the following
\begin{thm}
In the nonlocal gravity (\ref{actionLambda})-\Eq{gaga}-\Eq{gadel} in vacuum {\rm (}${\bf T}=0${\rm )}, the gravitational perturbations of maximally symmetric spacetimes satisfy the same EOM of the perturbations in Einstein gravity with a cosmological constant in vacuum.
\end{thm}
The proof of Theorem 2 proceeds as that of Theorem 1 in Sec.\ \ref{Nonlocal Gravity without cosmological constant}. The difference is that now we expand the metric around a maximally symmetric solution  $g^{(0)}_{\mu\nu}$ of $E_{\mu\nu}(g^{(0)}) = 0$, i.e., around a solution of Einstein equations in vacuum with a cosmological constant. By means of the expansion (\ref{hEXP}), we expand $E_{\mu\nu}(g_{\mu\nu})$ in powers of $\epsilon$ as
\be
E_{\mu\nu}(g_{\mu\nu}) = \sum_{n=0}^\infty \epsilon^n E^{(n)}_{\mu\nu} \, ,
\ee
where $E^{(0)}_{\mu\nu} = E_{\mu\nu}(g^0_{\mu\nu}) = 0$. Expanding the right-hand side of (\ref{EOMLF}) in powers of $\epsilon$, one obtains a recursive relation identical to Eq.\ (\ref{Recursive equations G}), which can be used to show that $E^{(0)}_{\mu\nu}=0$ implies $E^{(n)}_{\mu\nu}=0$ for any $n > 0$. Thus,
\be
{\bf E}^{{\rm NL}\,(n)} = 0 \quad \Longrightarrow \quad {\bf E}^{(n)}=0 \, ,
\ee
where ${\bf E}^{{\rm NL}\,(n)}$ is the perturbative expansion of the nonlocal EOM (\ref{EOML}), namely, 
\be
E^{\rm NL}_{\mu\nu}(g_{\mu\nu}) = \sum_{n=0}^\infty \epsilon^n E^{{\rm NL}\,(n)}_{\mu\nu} \, .
\ee

The stability of maximally symmetric spacetimes follows from Theorem 2. Therefore,
\begin{cor}
If a maximally symmetric spacetime is stable in Einstein gravity, it is also stable in NLG (\ref{actionLambda})-\Eq{gaga}-\Eq{gadel}. 
\end{cor}
All maximally symmetric spacetimes are stable at all perturbative orders in nonlocal gravity if they are stable in Einstein gravity in the presence of a cosmological constant. 

\section{Summary and applications} \label{concl}

We have proved that, in order to show the nonlinear stability of a Ricci-flat or a maximally symmetric spacetime at any perturbative order in nonlocal gravity, one needs to address the same issue in Einstein gravity. The stability equations of the Ricci-flat solutions of the nonlocal theory have shown to be exactly the same of general relativity. Therefore, if spacetime is stable in Einstein gravity, then it is stable in nonlocal gravity as well. The major advantage of this result is that we already know the stability properties of Ricci-flat solutions in general relativity and we do not need to repeat calculations such as those of Ref.\ \cite{WR}. One of the consequences is that Minkowski spacetime is stable in nonlocal gravity to all perturbative orders in the gravitational perturbation, as found previously in \cite{BrMo}. Furthermore, the extra degrees of freedom described by a symmetric tensor field never propagate on the Minkowski background.

Finally, Ricci-flat and maximally symmetric spacetimes are stable to all perturbative orders in the gravitational perturbation if they are stable in Einstein local gravity. The tensorial field describing the extra degree of freedom propagates neither on Ricci flat nor on maximally symmetric spacetimes. This paper confirms and extends the perturbative results found in \cite{Dona:2015tra, Koshelev:2017ebj}.

An interesting application of these results could be in the hot topic of gravitational-wave astronomy. In general relativity (see, e.g., \cite{Mag07} for an overview), the generation of GWs has been considered in models of astrophysical objects such as black-hole and neutron-star systems, while the propagation of GWs has been considered both on Minkowski spacetime (approximately representing the space neither too far nor too near a compact object) and on a Friedmann--Lema\^itre--Robertson--Walker (FLRW) cosmological background, where GWs propagate at large distances and are eventually observed on Earth. Both the production and propagation of GWs have not yet received attention in the context of NLG. We suggest to direct the effort to the study of the \emph{production} of GWs, since, as we are going to argue, nonlocality does not affect their \emph{propagation} appreciably.

Consider first general relativity in four dimensions. The linearized propagation equation of GWs on a FLRW background is $\B h=0$, where $h(t,{\bf x})$ is the amplitude of either tensor polarization mode \cite{MFB}. From the solution of this equation, or from a fairly simple scaling argument, one can recast the GW amplitude in terms of the redshift $1+z=1/a$ ($a$ is the FLRW scale factor) and a physical observable, the luminosity distance $d_L^\textsc{gw}=(1+z)\int_0^z\rmd z/H$ of the source, where $H=\dot a/a$ [not to be confused with the ${\rm H}$ in the form factor \Eq{LicForm}] is the Hubble parameter: $h\propto (d_L^\textsc{gw})^{-1}$ \cite{Mag07}. 
This information is sufficient to see what happens in NLG. At the linear level, the EOM \Eq{LastEOM} and \Eq{EOMLF} are valid also when the Lichnerowicz operator is replaced by the d'Alembertian $\B$. Ignoring the small contribution of $\Lambda$, one ends up with the linearized perturbation equation \cite{modesto,KS3}
\be
\B\tilde h=0\,,\qquad \tilde h=\rme^{{\rm H}} h\,.
\ee
Using the same scaling argument as in general relativity with $h$ replaced by $\tilde h$, for entire form factors we have
\be
\tilde h\propto \frac{1}{d_L^\textsc{gw}}\qquad\Rightarrow\qquad h\propto \rme^{-{\rm H}} \frac{1}{d_L^\textsc{gw}}\,.
\ee
We can estimate the nonlocal correction in the right-hand side for the string-related form factor ${\rm H}=-\ell_*^2\B=\ell_*^2(\p_t^2+3H\p_t)$ in the homogeneous approximation and, crudely, an approximately constant $H\simeq H_0$, so that $z\simeq\rme^{-H_0 (t-t_0)}-1$ and $d_L\simeq (z+1)z/H_0\simeq [\rme^{-2H_0(t-t_0)}-\rme^{-H_0(t-t_0)}]/H_0$. Since $\rme^{-{\rm H}}\rme^{n H_0t}=\rme^{-n(n+3)(\ell_*H_0)^2}\rme^{n H_0t}$, at large redshift $h\sim H_0\rme^{-10(\ell_*H_0)^2}\rme^{2H_0(t-t_0)}$, while at small redshift $z\simeq -H_0 (t-t_0)\ll 1$ one has $h\sim H_0\rme^{-3(\ell_*H_0)^2}\rme^{2H_0(t-t_0)}$. Overall, $h\sim \rme^{-c (\ell_*H_0)^2}/d_L^\textsc{gw}$, where $c=O(1)\!-\!O(10)$.

Are these nonlocality effects measurable? To answer this question, we may look into standard sirens, which are sources both of GWs and light. The binary neutron star merger GW170817 is the first known example \cite{Ab17b}. If propagation of electromagnetic waves were affected in the same way by the form factor, then the ratio between the luminosity distance $d_L^\textsc{gw}$ measured by an interferometer and the luminosity distance $d_L^\textsc{em}$ measured for the optical counterpart would be equal to 1. However, if light is not affected by nonlocality, we have
\be\label{dd}
\frac{d_L^\textsc{gw}}{d_L^\textsc{em}}\simeq 1+c (\ell_*H_0)^2\,,\qquad c=O(1)\!-\!O(10)\,,
\ee
and for $\ell_*=\lp$, the right-hand side would be of the order of $1+10^{-120}$, an effect completely unobservable compared with the estimated error $\De d_L/d_L\sim 0.001\!-\!0.1$ of present and future interferometers \cite{DHHJ,NHHDS,CaNi,Tam16}. For a power-law expansion $a=(t/t_0)^p$, $d_L\propto (t_0/t)^{2p}(t_0-t)$ and one can show that, again, the correction in the ratio \Eq{dd} is of the order of $(\ell_*/t_0)^2\sim 10^{-120}$. Increasing $\ell_*$ to particle-physics scales does not magnify this correction enough, since it is governed by the cosmological scale $H_0^{-1}\sim t_0\sim 10^{17}\,{\rm s}$.

The present paper sets a first step towards placing this argument, as well as future studies on GW production, on a rigorous ground. We have shown that metric perturbations (in particular, GWs) are well defined to all perturbative orders in Ricci-flat and maximally symmetric spacetimes where they are stable in general relativity, such as Minkowski and (anti--)de Sitter spacetime (see \cite{Bizon:2011gg,Craps:2014vaa,Craps:2015jma,Bizon:2015pfa} about anti--de Sitter (in)stability). These three situations cover most of the ranges of interest for the physics of astrophysical gravitational waves propagating at cosmological scales. Of course, de Sitter spacetime is not equivalent to FLRW, except in the strong slow-roll regime of inflation. However, it gets reasonably close to a cosmological setting as to make one confident that also cosmological backgrounds such as those already employed in the nonlocal literature are stable, at least at first order. The next step, which we leave for the future, will be to study the stability of FLRW spacetimes in the presence of matter. If stability depended on the absence of singularities in the FLRW metric, this topic will also be relevant for the big-bang problem, which may be resolved in the theory \cite{BMS,BKM1,CMNi}.

\section*{Acknowledgments} 

G.C.\ and L.M.\ are supported by the I+D Grant No.~IS2017-86497-C2-2-P of the Spanish Ministry of Science, Innovation and Universities.

\appendix

\begin{widetext}

\section{Two proofs of Theorem 1}\label{app1}

\begin{proof}[Proof 1] 
Let $g^{(0)}_{\mu\nu}$ be a Ricci-flat metric, i.e., such that $R_{\mu\nu}(g^{(0)}) = 0$. Consider small perturbations of  $g^{(0)}_{\mu\nu}$, by expanding the perturbed metric tensor $g_{\mu\nu}$ in powers of a small parameter $\epsilon\ll 1$ as in Eq.\ \Eq{hEXP}.
We can expand the Einstein tensor and all the other metric-dependent quantities in Eq.\ (\ref{LastEOMG}) by means of Eq.~(\ref{hEXP}):
\be
{\bf G}(g_{\mu\nu}) = \sum_{n=0}^\infty \epsilon^n {\bf G}^{(n)} , \qquad
{\bf G}^{(0)} \equiv {\bf G}(g^{(0)}) = 0 \, ,
\label{Gpert}
\ee
and
\be
\rme^{-{\rm H}(\s\Delta_{\rm L})} \equiv S  = \sum_{n=0}^\infty \epsilon^n S^{(n)}\, ,  \qquad
\mathcal{Q}_2 = \sum_{n=0}^\infty \epsilon^n \mathcal{Q}_2^{(n)} \, .
\label{Expansion Exponential and Q2}
\ee
Equation (\ref{Gpert}) also implies that it must be ${\bf G} \sim \epsilon^m$ for some integer $m \geqslant 1$, since ${\bf G}^{(r)} =0 $ for $r < m$ and at least the first term ${\bf G}^{(0)}$ is zero. As a consequence of this fact, one has that the dominant contribution in $\mathcal{Q} ({\bf Ric})$ in Eq.\ (\ref{LastEOM}) will be given by the quadratic terms included in ${\bf G} \, \mathcal{Q}_2 {\bf G}$, and we are allowed to neglect all the  terms at least cubic in ${\bf G}$, as we have done in Eq.~(\ref{LastEOMG}).

Substituting Eqs.\ (\ref{Gpert}) and (\ref{Expansion Exponential and Q2}) into the EOM (\ref{LastEOM}) or (\ref{LastEOMG}), one has
\be
{\bf G}^{(n)} = \sum_{h=0}^n \sum_{k=0}^h \sum_{q=0}^k S^{(n-h)}  {\bf G}^{(h-k)}  \, \mathcal{Q}_2^{(k-q)} {\bf G}^{(q)}   \, .
\label{Recursive equations G}
\ee 
At this point, it is straightforward to show that the condition ${\bf G}^{(0)}= 0$ implies that all the terms ${\bf G}^{(n)}$ must be zero. For instance, at first order in $\epsilon$ Eq.\ (\ref{Recursive equations G}) gives 
\be
&& {\bf G}^{(1)} =  S^{(0)}  \left({\bf G}^{(1)}  \, \mathcal{Q}_2^{(0)} {\bf G}^{(0)} + {\bf G}^{(0)}  \, \mathcal{Q}_2^{(1)} {\bf G}^{(0)} + {\bf G}^{(0)}  \, \mathcal{Q}_2^{(0)} {\bf G}^{(1)}  \right) 
+S^{(1)}  {\bf G}^{(0)}  \, \mathcal{Q}_2^{(0)} {\bf G}^{(0)}
= 0 \, .\nonumber
\ee 
At second order, using ${\bf G}^{(0)}= {\bf G}^{(1)}= 0 $, Eq.\ (\ref{Recursive equations G}) gives ${\bf G}^{(2)}= 0 $, and proceeding by recursion one has that ${\bf G}^{(n)}= 0$ for any $n$. Therefore, 
\be
{\bf \mathcal{E}}^{(n)} = 0 \quad \Longrightarrow \quad 
{\bf G}^{(n)}  = 0  \, .
\label{EoM Perturbations2}
\ee
Equation (\ref{EoM Perturbations}) tells us that the perturbed solution (\ref{hEXP}) must be a solution of the Einstein equations in vacuum, which concludes the proof. 
\end{proof}
\begin{proof}[Proof 2] Consider the action
\be
S[g,\phi,\chi] &=& \frac{1}{2\kappa^2} \int \rmd^D x \sqrt{-g} \left[\vphantom{\frac12}R + 2 \, G_{\mu \nu}\,  \gamma(\Delta_{\rm L}) \, \phi^{\mu\nu} 
- \phi_{\mu\nu} \, \gamma(\Delta_{\rm L}) \, \phi^{\mu\nu} +R \, \gamma( \Delta_{\rm L} )\, \chi
+ \frac{1}{D-2} \chi \, \gamma(\Delta_{\rm L}) \, \chi \right] .\label{actionAUX}
\ee
The EOM for the scalar $\chi$ and the tensor $\phi_{\mu\nu}$ are easy to derive: 
\be
&& \frac{\delta S}{\delta \chi} = 0 \quad \Longrightarrow \quad \chi = G=-\frac{D-2}{2}R \, , \label{EOMaux}\\
&& \frac{\delta S}{\delta \phi^{\mu\nu} } = 0 \quad \Longrightarrow \quad \phi_{\mu\nu} = G_{\mu\nu} \,. 
\label{EOMAUX}
\ee
Eliminating the auxiliary fields from the action (\ref{actionAUX}) we end up with (\ref{action}) with the form factor (\ref{LicForm}). Notice that the on-shell equation (\ref{EOMAUX}) implies $\nabla^\mu \phi_{\mu\nu} =0$ and $\chi=\phi^\mu_\mu$.

If the form factor $\gamma(\Delta_{\rm L})={\rm const}+O(\B)$ has a nonvanishing constant term in its analytic expansion, then the solution (\ref{EOMAUX}) is unique. Otherwise, if the form factor $\gamma(\Delta_{\rm L})$ has a vanishing constant term, then the action (\ref{actionAUX}) is invariant under the symmetry 
\bs\label{GINV}\ba
\phi_{\mu\nu}^\prime(x) &=& \phi_{\mu\nu}(x) + f_{\mu\nu}^{(\phi)}(x)\,,\, \qquad \mbox{where} \qquad \Box f_{\mu\nu}^{(\phi)} (x) = 0 \,,\\ 
\chi^\prime(x) &=& \chi(x) + f^{(\chi)}(x) \,,\,\quad\qquad \mbox{where} \qquad \Box f^{(\chi)}(x) = 0\,.
\ea\es
Therefore, the most general solution will be:
\ba
\chi &=& G=-\frac{D-2}{2}R + \chi_0\,, \qquad \mbox{where} \qquad \Box \chi_0 = 0 , \\
\phi_{\mu\nu} &=& G_{\mu\nu} + \phi_{0, \mu\nu}\,,\,\,\,\qquad\qquad\quad \mbox{where} \qquad \Box \phi_{0, \mu\nu} = 0 \, .
\ea
However, $\chi_0$ and $\phi_{0, \mu\nu}$ can always be fixed to zero using the gauge freedom (\ref{GINV}). Therefore, Eqs.\ (\ref{EOMaux}) and (\ref{EOMAUX}) are the most general gauge-invariant solutions.
 
Explicitly computing the variation of the action (\ref{actionAUX}) with respect to the metric $g_{\mu\nu}$, we get
\be 
\hspace{-0.5cm}
\delta S(g,\phi,\chi) &=&\int \rmd^D x \sqrt{-g}  \Big\{G_{\mu\nu} \, \delta g^{\mu\nu} 
- \frac{1}{2} g_{\mu\nu} \Big[ 2 \, G_{\alpha \beta} \, \gamma(\Delta_{\rm L }) \, \phi^{\alpha \beta} 
- \phi_{\alpha\beta} \, \gamma(\Delta_{\rm L}) \, \phi^{\alpha\beta} +R \gamma(\Delta_{\rm L}) \chi\nonumber\\ &&
+ \frac{1}{D-2} \chi \, \gamma(\Delta_{\rm L}) \, \chi  \Big] \delta g^{\mu\nu}
 +2 \, \delta G_{\alpha \beta} \, \gamma(\Delta_{\rm L}) \phi^{\alpha \beta} - \delta(g_{\mu\alpha}g_{\nu\beta}) \phi^{\alpha\beta} \, \gamma(\Delta_{\rm L}) \, \phi^{\mu\nu} \nonumber \\
&& +\delta R \, \gamma(\Delta_{\rm L}) \, \chi
+2 \, G_{\alpha \beta} \, \delta \gamma(\Delta_{\rm L}) \, \phi^{\alpha \beta} - \phi_{\mu\nu} \, \delta \gamma(\Delta_{\rm L}) \, \phi^{\mu\nu}+R \, \delta \gamma(\Delta_{\rm L}) \, \chi
+ \frac{1}{D-2} \chi \, \delta \gamma(\Delta_{\rm L}) \, \chi \Big\} .\label{variationC}
\ee
Replacing (\ref{EOMaux}) and (\ref{EOMAUX}) into (\ref{variationC}),\footnote{Since (\ref{variationC}) is a variation and not the EOM, the replacement mentioned in the text must be done carefully, namely, it cannot be done in $\delta {\bf G}$ and $\delta R$. Moreover, in (\ref{EOMphi}), by $O(\phi^2)$ we also mean products $R_{\mu\nu} \phi^{\mu\nu}$ in which the EOM (\ref{EOMAUX}) has been replaced.} we end up with the following equation of motion for the field $\phi_{\mu\nu}$ up to operators quadratic in the same field,
\be
E_{\mu\nu}^{\phi} :=
\phi_{\mu\nu} + 2 \frac{\delta R_{\alpha \beta}}{\delta g^{\mu\nu}} \gamma(\Delta_{\rm L}) \phi^{\alpha \beta} + \mathcal{Q} ({\phi}^2) =0 \, . 
\label{EOMphi}
\ee
The background Ricci-flat solution is now given by the pair $(g^{(0)}_{\mu\nu}, \phi^{(0)})$ solving the EOM (\ref{EOMAUX}) and (\ref{EOMphi}):
\be
\quad \phi^{(0)}_{\mu\nu} = G_{\mu\nu}(g^{(0)}) = 0 .
\label{phig}
\ee
Since $\phi_{\mu\nu} = G_{\mu\nu}$ on shell, then $\nabla_{\alpha} \phi^{\alpha \beta} = 0$ (on shell) and we can use the results in \cite{CMM} to simplify the EOM (\ref{EOMphi}) and finally get
\be
\rme^{{\rm H}(\s\Delta_{\rm L})} {\bm \phi} =  \mathcal{Q}({\phi}^2) 
=   {\bm \phi} \mathcal{Q}_2 {\bm \phi} + \cdots  
\quad \Rightarrow \quad 
{\bm \phi} = \rme^{-{\rm H}(\s\Delta_{\rm L})} \mathcal{Q}({\phi}^2)= \rme^{-{\rm H}(\s\Delta_{\rm L})} \left({\bm \phi} \mathcal{Q}_2 {\bm \phi} + \cdots \right) \, ,
\label{EOMphi2}
\ee
where the dots stand for terms at least cubic in $\phi$. Using now the metric expansion (\ref{hEXP}) and a similar expansion for the tensor field $\phi_{\mu\nu}$ around its vacuum  $\phi^{(0)}_{\mu\nu}= 0$, 
\be
\phi_{\mu\nu} =   \sum_{n=0}^{\infty} \epsilon^n \phi^{(n)}_{\mu\nu} \, , 
\label{phiEXP}
\ee
we can solve simultaneously Eqs.\ (\ref{EOMAUX}) and (\ref{EOMphi2}) to all perturbative orders, obtaining
\be
\bm{\phi}^{(n)} = \sum_{h=0}^n \sum_{k=0}^h \sum_{q=0}^k S^{(n-h)}  \bm{\phi}^{(h-k)}  \, \mathcal{Q}_2^{(k-q)} \bm{\phi}^{(q)}   \, .
\label{Recursive equations phi}
\ee 
From Eq.\ (\ref{Recursive equations phi}), it is straightforward to show that $\bm{\phi}^{(0)} = 0$ implies $\bm{\phi}^{(n)} = 0$ for all $n \geqslant 0$. Therefore, the tensor field $\bm{\phi}$ is identically zero in vacuum, $\phi_{\mu\nu} = G_{\mu\nu} =0$, and all the solutions of the theory (\ref{actionAUX}) must satisfy the Einstein equations. In other words, the tensor field $\phi_{\mu\nu}$ is not a dynamical field on any Ricci-flat background at any perturbative order. 
%
\end{proof}

\section{Proof of the EOM \Eq{EOMLambda5}}\label{app2}

The variation of the Ricci tensor acting on a rank-2 tensor is
\be\label{varric}
\frac{\delta R_{\alpha\beta}}{\delta g^{\mu \nu}  }=\frac{1}{2}g_{\alpha(\mu}g_{\nu)\beta}\Box+\frac{1}{2}g_{\mu\nu}\nabla_\alpha
\nabla_\beta-g_{\alpha(\mu|}\nabla_\beta\nabla_{|\nu)} \, , 
\ee
which allows us to write explicitly the EOM \Eq{EOML} as
\be
E_{\mu\nu} + \Big[
 g_{\alpha(\mu} g_{\nu) \beta}  \Box + \underbrace{\vphantom{\frac12} g_{\mu\nu} \nabla_\alpha \nabla_\beta}_{\circled{1}}
- \underbrace{\vphantom{\frac12}\left( g_{\alpha \mu} \nabla_\beta \nabla_\nu + g_{\alpha \nu} \nabla_\beta \nabla_\mu \right)}_{\circled{2}}  - 2\Lambda \, g_{\mu \alpha} \, g_{\nu \beta} \Big] \gamma(\Delta_\Lambda) E^{\alpha \beta} 
+ O({\bf E^2}) = 0 
\,.
\label{EOMLambda}
\ee
The two terms $\circled{1}$ and $\circled{2}$ can be dealt with separately. We will show that
\be
\circled{1}&=& O({\bf E^2})\,,\label{useA}\\
\circled{2} &=& 2 R_{\mu \beta \nu \lambda} \, \gamma(\Delta_\Lambda) \, E^{\beta \lambda} - 2 \Lambda \, \gamma(\Delta_\Lambda) \, E_{\mu\nu}  +  O({\bf E^2}) \, . \label{2E}
\ee

\subsection{Proof of Eq.\ \Eq{useA}} \label{AppendixA}

Using the definition of the generalized Lichnerowicz operator (\ref{lichLambda}), we get 
\be
\hspace{-.1cm}\nabla^{\mu} \left[ \Delta_\Lambda \, \left(G_{\mu\nu} + \Lambda \, g_{\mu\nu} \right) \right] &=& 
- \nabla^\mu \left[ \Box  \left(G_{\mu\nu} + \Lambda \, g_{\mu\nu} \right) 
+ 2 R_{\mu\rho \nu \sigma} \left(G^{\rho \sigma} + \Lambda \, g^{\rho \sigma} \right)  \right] +
O({\bf E^2}) \nonumber \\
&=& 
- \Big[ \underbrace{\vphantom{\frac12}\nabla^\mu  \Box  \left(G_{\mu\nu} + \Lambda \, g_{\mu\nu} \right) }_{\circled{I}}
+ \underbrace{\vphantom{\frac12}2 \, \nabla^\mu \left( R_{\mu\rho \nu \sigma} \left(G^{\rho \sigma} + \Lambda \, g^{\rho \sigma} \right) \right) }_{\circled{II}}\Big] +
O({\bf E^2}), 
\label{BianchiLich-1}
\ee
because the contributions of the second and third tensor in the definition (\ref{lichLambda}) are quadratic in the EOM. 

Let us start with $\circled{I}$:
\be
 \circled{I} &=& \nabla^\mu  \nabla_\alpha \nabla^\alpha  \left(G_{\mu\nu} + \Lambda \, g_{\mu\nu} \right) 
=  \nabla_\alpha  \nabla^\mu  \nabla^\alpha  \left(G_{\mu\nu} + \Lambda \, g_{\mu\nu} \right)
+ [\nabla^\mu, \nabla_\alpha]  \nabla^\alpha  \left(G_{\mu\nu} + \Lambda \, g_{\mu\nu} \right) \nonumber \\
&=& \nabla_\alpha  \nabla^\alpha  \underbrace{\nabla^\mu   \left(G_{\mu\nu} + \Lambda \, g_{\mu\nu} \right)}_{= 0 \,\,\, {\mbox{(Bianchi id.)}}}
+  \underbrace{\nabla_\alpha  [ \nabla^\mu,  \nabla^\alpha]   \left(G_{\mu\nu} + \Lambda \, g_{\mu\nu} \right)}_{\circled{a}}
+ \underbrace{[\nabla^\mu, \nabla_\alpha]  \nabla^\alpha  \left(G_{\mu\nu} + \Lambda \, g_{\mu\nu} \right) }_{\circled{b}}. 
\label{CI}
\ee
Using the short notation defined in (\ref{EinsteinLambda}) and the commutator of covariant derivatives on a symmetric tensor
\be
[\nabla^\mu, \nabla_\alpha] \, X_{\mu \nu} =
R_{\nu \lambda \mu \alpha} \, X^{\lambda\mu} + R_{\lambda\alpha}\, X^{\lambda}\,_{\nu} \, ,
\label{commutatori2}
\ee
we can write $\circled{a}$ as
\be
\circled{a} &=& \nabla^\alpha  [ \nabla^\mu,  \nabla_\alpha]  E_{\mu\nu} 
=  \nabla^\alpha  \left[ - R^{\lambda}\,_{\nu \mu \alpha} \, E^{\mu}\,_{\lambda} + R_{\lambda\alpha}\, E^{\lambda}\,_{\nu} \right]\nonumber \\
&=& ( \nabla^\alpha R_{\lambda\alpha} ) \, E^{\lambda}\,_{\nu}
+  R_{\lambda\alpha} ( \nabla^\alpha E^{\lambda}\,_{\nu} ) 
- ( \nabla^\alpha  R^{\lambda}\,_{\nu \mu \alpha}) \, E^{\mu}\,_{\lambda} 
- R^{\lambda}\,_{\nu \mu \alpha} \, ( \nabla^\alpha E^{\mu}\,_{\lambda} ) \, . 
\label{a0}
\ee
We can replace the Ricci tensor in the first term with $R_{\lambda \alpha} - \Lambda g_{\lambda \alpha}$ thanks to the metric compatibility condition $\nabla^\alpha  g_{\lambda \alpha}=0$.
Moreover, we can use the contracted Bianchi identities to simplify the third term. Starting from 
\be
- \nabla^\alpha  R_{\nu \sigma \beta \alpha} = \nabla_\nu R_{\sigma \beta} - \nabla_\sigma R_{\nu \beta} \, , 
\label{B1}
\ee
and replacing again the Ricci tensor with $R_{\lambda \alpha} - \Lambda g_{\lambda \alpha}$ in (\ref{B1}), we end up with 
\be
- \nabla^\alpha  R_{\nu \sigma \beta \alpha} = \nabla_\nu (R_{\sigma \beta}  - \Lambda b_{\sigma \beta}) - \nabla_\sigma (R_{\nu \beta} - \Lambda g_{\nu \beta})  \, .
\label{B2L}
\ee
Therefore, the third term in (\ref{a0}) is $O(E^2)$ (quadratic in the EOM) and, going back to (\ref{a0}), $\circled{a}$ simplifies to 
\be
&& \circled{a}  = \underbrace{( \nabla^\alpha R_{\lambda\alpha} ) \, E^{\lambda}\,_{\nu}}_{O(\bf E^2)} 
+  R_{\lambda\alpha} ( \nabla^\alpha E^{\lambda}\,_{\nu} ) 
- \underbrace{( \nabla^\alpha  R^{\lambda}\,_{\nu \mu \alpha}) \, E^{\mu}\,_{\lambda} }_{O({\bf E^2})}
- R^{\lambda}\,_{\nu \mu \alpha} \, ( \nabla^\alpha E^{\mu}\,_{\lambda} ) \, . 
\label{a0b}
\ee
Looking at the second term in (\ref{a0b}), it turns out to be equivalent to 
\be
R_{\lambda\alpha}  ( \nabla^\alpha E^{\lambda}\,_{\nu} ) =
(R_{\lambda\alpha} - \Lambda g_{\lambda \alpha} ) ( \nabla^\alpha E^{\lambda}\,_{\nu} ) =
O({\bf E^2}) ,
\label{B23}
\ee
since the term proportional to $\Lambda$ vanishes as a consequence of $\nabla^\alpha E_{\alpha \nu} = 0$. Finally,
\be
\circled{a}  &=& \underbrace{( \nabla^\alpha R_{\lambda\alpha} ) \, E^{\lambda}\,_{\nu}}_{O(\bf E^2)} 
+  \underbrace{R_{\lambda\alpha} ( \nabla^\alpha E^{\lambda}\,_{\nu} ) }_{O({\bf E^2})}
- \underbrace{( \nabla^\alpha  R^{\lambda}\,_{\nu \mu \alpha}) \, E^{\mu}\,_{\lambda} }_{O({\bf E^2})}
- R^{\lambda}\,_{\nu \mu \alpha} \, ( \nabla^\alpha E^{\mu}\,_{\lambda} ) \nonumber\\
&=& - R^{\lambda}\,_{\nu \mu \alpha} \, ( \nabla^\alpha E^{\mu}\,_{\lambda} ) + O({\bf E^2})\,.\label{a0c}
\ee
In order to simplify $\circled{b}$, we need the commutator of two covariant derivatives acting on a rank-3 tensor,
\be
[\nabla_\rho,\nabla_{\mu_1}]X^{\mu_1\mu_2\mu_3} = R^{\mu_1}_{\ \lambda \rho \mu_1} X^{\lambda \mu_2 \mu_3}
+ R^{\mu_2}_{\ \lambda \rho \mu_1} X^{\mu_1 \lambda \mu_3} 
+ R^{\mu_3}_{\ \lambda \rho \mu_1} X^{\mu_1 \mu_2 \lambda}\,.\label{perb2}
\ee
Using the EOM $E_{\mu\nu}=0$, the $\circled{b}$-term reads 
\be
\circled{b} &=& g_{\delta \nu} [\nabla_\mu, \nabla_\alpha]  \nabla^\alpha  E^{\mu \delta} = 
g_{\delta \nu} R^\alpha\,_{\lambda \mu \alpha} \nabla^\lambda E^{\mu \delta} 
+g_{\delta \nu} R^\mu\,_{\lambda \mu \alpha} \nabla^\alpha E^{\lambda \delta} 
+g_{\delta \nu} R^\delta\,_{\lambda \mu \alpha} \nabla^\alpha E^{\mu \lambda } \nonumber \\
&=& - g_{\delta \nu} R_{\lambda \mu} \nabla^\lambda E^{\mu \delta} 
+  g_{\delta \nu} R_{\lambda \alpha} \nabla^\alpha E^{\lambda \delta} 
+ R_{\nu \lambda \mu \alpha} \nabla^\alpha E^{\mu \lambda } = 
- R_{\alpha \mu \nu \lambda} \nabla^\alpha E^{\mu \lambda }\label{bF} \, .
\ee
Therefore, plugging (\ref{a0c}) and  (\ref{bF}) into (\ref{CI}) we find 
\be
\circled{I} = -2  R_{\alpha \mu \nu \lambda} \nabla^\alpha E^{\mu \lambda } \, .
\label{IF}
\ee

We now move on to calculate $\circled{II}$:
\be
\circled{II} = 2 \, \nabla^\mu \left( R_{\mu\rho \nu \sigma} E^{\rho \sigma} \right) = 
2 \, \underbrace{\left( \nabla^\mu R_{\mu\rho \nu \sigma} \right) E^{\rho \sigma}}_{O({\bf E ^2}), \,\,\, \mbox{see (\ref{B2L})}}  +
2  \, R_{\mu\rho \nu \sigma} \left( \nabla^\mu E^{\rho \sigma}  \right) = 
2  \, R_{\mu\rho \nu \sigma} \left( \nabla^\mu E^{\rho \sigma}  \right)  + O({\bf E^2}) \, . 
\label{IIF}
\ee
Replacing (\ref{IF}) and (\ref{IIF}) into (\ref{BianchiLich-1}), we get
\be 
 \nabla^{\mu} \left( \Delta_\Lambda \, E_{\mu\nu} \right) = O({\bf E^2}) \, . 
 \label{GeneralDiv}
\ee
Reiterating the procedure that led us to (\ref{GeneralDiv}) one can see that
\be
\nabla^\mu \left( \Delta_\Lambda \Delta_\Lambda \cdots \Delta_\Lambda E_{\mu\nu} \right)  = O({\bf E^2})\,,
\ee
and for an analytic form factor $\gamma(\Delta_\Lambda)$, we finally get the identity
\be
\nabla^{\mu} \left[\gamma(\Delta_\Lambda) E_{\mu\nu}\right] = O({\bf E^2}) \,,
\label{BianchiLich}
\ee
thanks to which we conclude that \Eq{useA} holds.

\subsection{Proof of Eq.\ \Eq{2E}} \label{AppendixB}

The term $\circled{2}$ can be manipulated as follows:
\be
\circled{2} &=& - \left( g_{\alpha \mu} \nabla_\beta \nabla_\nu + g_{\alpha \nu} \nabla_\beta \nabla_\mu \right)\gamma(\Delta_\Lambda) \, E^{\alpha \beta}\nonumber\\
&=& -\left( g_{\alpha \mu} \nabla_\nu \nabla_\beta + g_{\alpha \mu} [ \nabla_\beta, \nabla_\nu ]
+ g_{\alpha \nu} \nabla_\mu \nabla_\beta + g_{\alpha \nu} [ \nabla_\beta , \nabla_\mu] 
 \right) \gamma(\Delta_\Lambda) \, E^{\alpha \beta} \, . \label{2b}
\ee
The first and the third term in (\ref{2b}) are both zero because of (\ref{BianchiLich}), and using
\be
 [\nabla_\beta, \nabla_\mu] \, X^{\alpha \beta} =
R^\alpha\,_{\lambda \beta \mu} \, X^{\lambda \beta} + R_{\lambda\mu}\, X^{\lambda \alpha} \, , 
\label{commutatori}
\ee
we can replace the commutators in $\circled{2}$ with the Riemann and Ricci tensor, 
\be
\hspace{-1cm}\circled{2} &=& - \left( g_{\alpha \mu} [ \nabla_\beta, \nabla_\nu ]+ g_{\alpha \nu} [ \nabla_\beta , \nabla_\mu] 
 \right)  \gamma(\Delta_\Lambda) \, E^{\alpha \beta} +  O({\bf E^2}) 
 \nonumber \\
\hspace{-1cm} &=& - \left[ g_{\alpha \mu} R^\alpha\,_{\lambda \beta \nu} \gamma(\Delta_\Lambda) \, E^{\lambda \beta}
 + g_{\alpha \mu} R_{\lambda  \nu} \gamma(\Delta_\Lambda) \, E^{\lambda \alpha}
 + g_{\alpha \nu} R^\alpha\,_{\lambda \beta \mu} \gamma(\Delta_\Lambda) \, E^{\lambda \beta}
 + g_{\alpha \nu} R_{\lambda  \mu} \gamma(\Delta_\Lambda) \, E^{\lambda \alpha}\right]
 +  O({\bf E^2})\,. \label{2c}
\ee
Up to a reshuffling of the indices, the two operators with the Riemann tensor in (\ref{2c}) are identical, so that
\be
\circled{2} 
 = 2 R_{\mu \lambda \nu \beta} \, \gamma(\Delta_\Lambda) \, E^{\lambda \beta}
-  R_{\lambda  \nu} \, \gamma(\Delta_\Lambda) \, E^{\lambda}\,_\mu 
-  R_{\lambda  \mu} \, \gamma(\Delta_\Lambda) \, E^{\lambda}\,_\nu 
  +  O({\bf E^2}) 
  \, .
 \label{2d}
\ee
We now add and subtract $\Lambda g_{\lambda \nu} \gamma(\Delta_\Lambda) \, E^{\lambda}\,_\mu$ and $\Lambda g_{\lambda \mu}  \, \gamma(\Delta_\Lambda) \, E^{\lambda}\,_\nu$ in order to replace the Ricci tensor with the tensor (\ref{EinsteinLambda}):
\be
\circled{2} 
 &=& 2 R_{\mu \lambda \nu \beta } \gamma(\Delta_\Lambda) \, E^{\lambda \beta} 
 -  R_{\lambda  \nu} \, \gamma(\Delta_\Lambda) \, E^{\lambda}\,_\mu 
-  R_{\lambda  \mu} \, \gamma(\Delta_\Lambda) \, E^{\lambda}\,_\nu \nonumber \\
&& + \Lambda g_{\lambda \nu} \gamma(\Delta_\Lambda) \, E^{\lambda}\,_\mu 
- \Lambda g_{\lambda \nu} \gamma(\Delta_\Lambda) \, E^{\lambda}\,_\mu
+\Lambda g_{\lambda \mu}  \, \gamma(\Delta_\Lambda) \, E^{\lambda}\,_\nu 
 - \Lambda g_{\lambda \mu}  \, \gamma(\Delta_\Lambda) \, E^{\lambda}\,_\nu 
  +  O({\bf E^2}) \nonumber \\
&=& 2 R_{\mu \lambda \nu \beta} \, \gamma(\Delta_\Lambda) \, E^{\lambda \beta}
-  \Lambda g_{\lambda  \nu} \, \gamma(\Delta_\Lambda) \, E^{\lambda}\,_\mu 
-  \Lambda g_{\lambda  \mu} \, \gamma(\Delta_\Lambda) \, E^{\lambda}\,_\nu + O({\bf E^2}),
\ee
yielding \Eq{2E}.

\subsection{Equations of motion}

Finally, from (\ref{2E}) and the identity (\ref{BianchiLich}), the EOM (\ref{EOMLambda}) turn into
\be
&& E_{\mu\nu} +  \left( \Box - 2 \Lambda \right) \gamma(\Delta_\Lambda) E_{\mu \nu}
+ 2 R_{\mu \alpha \nu \beta} \, \gamma(\Delta_\Lambda) \, E^{\alpha \beta} - 2 \Lambda \, \gamma(\Delta_\Lambda) \, E_{\mu\nu} 
+ O({\bf E^2})= 0 \, , \nonumber \\
&& E_{\mu\nu} +  \left( \Box - 4 \Lambda \right) \gamma(\Delta_\Lambda) E_{\mu\nu}
+ 2 R_{\mu \alpha \nu \beta} \, \gamma(\Delta_\Lambda) \, E^{\alpha \beta } + O({\bf E^2})= 0 \, .
\label{EOMLambda2}
\ee
Now we add and subtract $(R_{\mu \alpha} - \Lambda\, g_{\mu\alpha}) \gamma(\Delta_\Lambda) E^{\alpha}\,_\nu$ and $(R_{\nu \alpha} - \Lambda \, g_{\nu\alpha}) \gamma(\Delta_\Lambda) E^{\alpha}\,_\mu$ to the EOM  (\ref{EOMLambda2}) in order to reconstruct the generalized Lichnerowicz operator $\Delta_\Lambda$ defined in (\ref{lichLambda}):
\be
&& E_{\mu\nu} +  \left( \Box - 4 \Lambda \right) \gamma(\Delta_\Lambda) E_{\mu\nu}
+ 2 R_{\mu \alpha \nu \beta} \, \gamma(\Delta_\Lambda) \, E^{\alpha \beta } \nonumber \\
&& + (R_{\mu \alpha} - \Lambda\, g_{\mu\alpha}) \gamma(\Delta_\Lambda) E^{\alpha}\,_\nu
- (R_{\mu \alpha} - \Lambda\, g_{\mu\alpha}) \gamma(\Delta_\Lambda) E^{\alpha}\,_\nu \nonumber \\
&& + (R_{\nu \alpha} - \Lambda \, g_{\nu\alpha}) \gamma(\Delta_\Lambda) E^{\alpha}\,_\mu
- (R_{\nu \alpha} - \Lambda \, g_{\nu\alpha} ) \gamma(\Delta_\Lambda) E^{\alpha}\,_\mu
+ O({\bf E^2})= 0 \, .
\label{EOMLambda3}
\ee
Up to $O({\bf E^2})$ terms, we get
\be
&& E_{\mu\nu}  \underline{ + \Box \, \gamma(\Delta_\Lambda) E_{\mu\nu}}- 4 \Lambda  \gamma(\Delta_\Lambda) E_{\mu\nu}
 \underline{+ 2 R_{\mu \alpha \nu \beta} \, \gamma(\Delta_\Lambda) \, E^{\alpha \beta }} \nonumber \\
&& 
\underline{- (R_{\mu \alpha} - \Lambda\, g_{\mu\alpha}) \gamma(\Delta_\Lambda) E^{\alpha}\,_\nu 
- (R_{\nu \alpha} - \Lambda \, g_{\nu\alpha} ) \gamma(\Delta_\Lambda) E^{\alpha}\,_\mu}
+ O({\bf E^2})= 0 \, .
\label{EOMLambda4}
\ee
The underlined operators reconstruct the Lichnerowicz operator $-\Delta_\Lambda$ and we end up with Eq.\ \Eq{EOMLambda5}.

\end{widetext}



\begin{thebibliography}{99}
\bibitem{Ab16a} B.P.\ Abbott {et al.} [LIGO Scientific and Virgo Collaborations], \tia{Observation of gravitational waves from a binary black hole merger} \doinn{10.1103/PhysRevLett.116.061102}{Phys.\ Rev.\ Lett.}{116}{061102}{2016} [\arX{1602.03837}].
\bibitem{Ab17b} B.P.\ Abbott {et al.} [LIGO Scientific and Virgo and Fermi-GBM and INTEGRAL Collaborations], \tia{Gravitational waves and gamma-rays from a binary neutron star merger: GW170817 and GRB 170817A} \doinn{10.3847/2041-8213/aa920c}{Astrophys.\ J.}{848}{L13}{2017} [\arX{1710.05834}].
\bibitem{P18I}  Y.\ Akrami {et al.} [Planck Collaboration], \tia{Planck 2018 results. I. Overview and the cosmological legacy of Planck} \arX{1807.06205}.
\bibitem{modesto} L.\ Modesto, \tia{Super-renormalizable quantum gravity} \doin{10.1103/PhysRevD.86.044005}{Phys.\ Rev.}{D}{86}{044005}{2012} [\arX{1107.2403}].
\bibitem{BCKM}  T.\ Biswas, A.\ Conroy, A.S.\ Koshelev, and A.\ Mazumdar, \tia{Generalized ghost-free quadratic curvature gravity} \doinn{10.1088/0264-9381/31/1/015022}{Classical Quantum Gravity}{31}{015022}{2014} [\doinn{10.1088/0264-9381/31/15/159501}{Erratum ibid.}{31}{159501}{2014}] [\arX{1308.2319}].
\bibitem{CaMo2} G.\ Calcagni and L.\ Modesto, \tia{Nonlocal quantum gravity and M-theory} \doin{10.1103/PhysRevD.91.124059}{Phys.\ Rev.}{D}{91}{124059}{2015} [\arX{1404.2137}].
\bibitem{modestoLeslaw} L.\ Modesto and L.\ Rachwa\l, \tia{Super-renormalizable and finite gravitational theories} \doin{10.1016/j.nuclphysb.2014.10.015}{Nucl.\ Phys.}{B}{889}{228}{2014} [\arX{1407.8036}].
\bibitem{universality} L.\ Modesto and L.\ Rachwa\l, \tia{Universally finite gravitational and gauge theories} \doin{10.1016/j.nuclphysb.2015.09.006}{Nucl.\ Phys.}{B}{900}{147}{2015} [\arX{1503.00261}].
\bibitem{Review} L.\ Modesto and L.\ Rachwa\l, \tia{Nonlocal quantum gravity: a review} \doin{10.1142/S0218271817300208}{Int.\ J.\ Mod.\ Phys.}{D}{26}{1730020}{2017}.
\bibitem{Buoninfante:2018lnh} L.~Buoninfante, G.~Lambiase, and M.~Yamaguchi, \tia{Nonlocal generalization of Galilean theories and gravity} \arX{1812.10105}.
\bibitem{Ste77} K.S.\ Stelle, \tia{Renormalization of higher-derivative quantum gravity} \doin{10.1103/PhysRevD.16.953}{Phys.\ Rev.}{D}{16}{953}{1977}.
\bibitem{Ste78} K.S.\ Stelle, \tia{Classical gravity with higher derivatives} \doinn{10.1007/BF00760427}{Gen.\ Rel.\ Grav.}{9}{353}{1978}.
\bibitem{Krasnikov} N.V.\ Krasnikov, \tia{Nonlocal gauge theories} \doinn{10.1007/BF01017588}{Theor.\ Math.\ Phys.}{73}{1184}{1987} [\ndoinn{http://www.mathnet.ru/php/archive.phtml?wshow=paper&jrnid=tmf&paperid=5624&option_lang=eng}{Teor.\ Mat.\ Fiz.}{73}{235}{1987}].
\bibitem{kuzmin} Yu.V.\ Kuz'min, \tia{The convergent nonlocal gravitation} Yad.\ Fiz.\ {50} (1989) 1630 [Sov.\ J.\ Nucl.\ Phys.\ {50} (1989) 1011].
\bibitem{Tom97} E.T.\ Tomboulis, \tia{Super-renormalizable gauge and gravitational theories} \oarX{hep-th/9702146}.
\bibitem{Sie03} W.\ Siegel, \tia{Stringy gravity at short distances} \oarX{hep-th/0309093}.
\bibitem{Kho06} J.\ Khoury, \tia{Fading gravity and self-inflation} \doin{10.1103/PhysRevD.76.123513}{Phys.\ Rev.}{D}{76}{123513}{2007} [\oarX{hep-th/0612052}].
\bibitem{PiSe} R.\ Pius and A.\ Sen, \tia{Cutkosky rules for superstring field theory} \doij{10.1007/JHEP10(2016)024}{J.\ High Energy Phys.}{1610}{024}{2016}; \doij{10.1007/JHEP09(2018)122}{Erratum-ibid.}{1809}{122}{2018} [\arX{1604.01783}].
\bibitem{brisceseUnitarity} F.\ Briscese and L.\ Modesto, \tia{Cutkosky rules and perturbative unitarity in Euclidean nonlocal quantum field theories} \arX{1803.08827}.
\bibitem{ChTo} P.\ Chin and E.T.\ Tomboulis, \tia{Nonlocal vertices and analyticity: Landau equations and general Cutkosky rule} \doij{10.1007/JHEP06(2018)014}{J.\ High Energy Phys.}{1806}{014}{2018} [\arX{1803.08899}].
\bibitem{shapiromodesto}  L.\ Modesto and I.L.\ Shapiro, \tia{Superrenormalizable quantum gravity with complex ghosts} \doin{10.1016/j.physletb.2016.02.021}{Phys.\ Lett.}{B}{755}{279}{2016} [\arX{1512.07600}]. 
\bibitem{LWqg}  L.\ Modesto, \tia{Super-renormalizable or finite Lee--Wick quantum gravity} \doin{10.1016/j.nuclphysb.2016.06.004}{Nucl.\ Phys.}{B}{909}{584}{2016}  [\arX{1602.02421}].
\bibitem{shapiro3} M.\ Asorey, J.L.\ L\'opez, and I.L.\ Shapiro,  \tia{Some remarks on high derivative quantum gravity} \doin{10.1142/S0217751X97002991}{Int.\ J.\ Mod.\ Phys.}{A}{12}{5711}{1997} [\oarX{hep-th/9610006}]. 
\bibitem{yaudong} Y.D.\ Li, L.\ Modesto, and L.\ Rachwa\l, \tia{Exact solutions and spacetime singularities in nonlocal gravity} \doij{10.1007/JHEP12(2015)173}{J.\ High Energy Phys.}{1512}{173}{2015} [\arX{1506.08619}].    
\bibitem{CM} G.\ Calcagni and L.\ Modesto, \tia{Stability of Schwarzschild singularity in non-local gravity} \doin{10.1016/j.physletb.2017.09.018}{Phys.\ Lett.}{B}{773}{596}{2017} [\arX{1707.01119}].
\bibitem{CMM} G.\ Calcagni, L.\ Modesto, and Y.S.\ Myung, \tia{Black-hole stability in non-local gravity} \doin{10.1016/j.physletb.2018.06.041}{Phys.\ Lett.}{B}{783}{19}{2018} [\arX{1803.08388}].
\bibitem{WR} T.\ Regge and J.A.\ Wheeler, \tia{Stability of a Schwarzschild singularity} \doinn{10.1103/PhysRev.108.1063}{Phys.\ Rev.}{108}{1063}{1957}.
\bibitem{Tse95} A.A.\ Tseytlin, \tia{On singularities of spherically symmetric backgrounds in string theory} \doin{10.1016/0370-2693(95)01228-7}{Phys.\ Lett.}{B}{363}{223}{1995} [\oarX{hep-th/9509050}].
\bibitem{FZdP}  V.P.\ Frolov, A.\ Zelnikov, and T.\ de Paula Netto, \tia{Spherical collapse of small masses in the ghost-free gravity} \doij{10.1007/JHEP06(2015)107}{J.\ High Energy Phys.}{1506}{107}{2015} [\arX{1504.00412}].
\bibitem{Fro15} V.P.\ Frolov, \tia{Mass gap for black-hole formation in higher-derivative and ghost-free gravity} \doinn{10.1103/PhysRevLett.115.051102}{Phys.\ Rev.\ Lett.}{115}{051102}{2015} [\arX{1505.00492}].
\bibitem{Fro16} V.P.\ Frolov, \tia{Notes on nonsingular models of black holes} \doin{10.1103/PhysRevD.94.104056}{Phys.\ Rev.}{D}{94}{104056}{2016} [\arX{1609.01758}].
\bibitem{Gia1} B.L.\ Giacchini and T.\ de Paula Netto, \tia{Weak-field limit and regular solutions in polynomial higher-derivative gravities} \doin{10.1140/epjc/s10052-019-6727-2}{Eur.\ Phys.\ J}{C}{79}{217}{2019} [\arX{1806.05664}].
\bibitem{Gia2} B.L.\ Giacchini and T.\ de Paula Netto, \tia{Effective delta sources and regularity in higher-derivative and ghost-free gravity} \arX{1809.05907}.
\bibitem{BrMo}  F.\ Briscese and L.\ Modesto, \tia{Nonlinear stability of Minkowski spacetime in nonlocal gravity} \arX{1811.05117}.
\bibitem{ChKl}  D.~Christodoulou and S.~Klainerman, \tia{The nonlinear stability of the Minkowski metric in general relativity} \procsinm{Nonlinear Hyperbolic Problems}{C.\ Carasso, P.\ Charrier, B.\ Hanouzet, and J.-L.\ Joly}{Springer-Verlag}{Berlin}{1989}.
\bibitem{Chr91} D.~Christodoulou, \tia{The stability of Minkowski spacetime} \procsinm{Proceedings of the International Congress of Mathematicians, Kyoto 1990}{I.\ Satake}{Springer-Verlag}{Berlin}{1991}.
\bibitem{LiRo}  H.~Lindblad and I.~Rodnianski, \tia{The global stability of Minkowski space-time in harmonic gauge} \doinn{10.4007/annals.2010.171.1401}{Ann.\ Math.}{171}{1401}{2010} [\oarX{math/0411109}].
\bibitem{Lin17} H.~Lindblad, \tia{On the asymptotic behavior of solutions to the Einstein vacuum equations in wave coordinates} \doinn{10.1007/s00220-017-2876-z}{Commun.\ Math.\ Phys.}{353}{135}{2017}. 
\bibitem{Briscese:2017vff} F.~Briscese and P.M.~Santini, \tia{On the occurrence of gauge-dependent secularities in nonlinear gravitational waves} \doinn{10.1088/1361-6382/aa7451}{Classical Quantum Gravity}{34}{144001}{2017} [\arX{1705.10990}].
\bibitem{BKM}   T.\ Biswas, A.S.\ Koshelev, and A.\ Mazumdar, \tia{Consistent higher derivative gravitational theories with stable de Sitter and anti--de Sitter backgrounds} \doin{10.1103/PhysRevD.95.043533}{Phys.\ Rev.}{D}{95}{043533}{2017} [\arX{1606.01250}].
\bibitem{KSML}  A.S.\ Koshelev, K.S.\ Kumar, L.\ Modesto, and L.\ Rachwa\l, \tia{Finite quantum gravity in dS and AdS spacetimes} \doin{10.1103/PhysRevD.98.046007}{Phys.\ Rev.}{D}{98}{046007}{2018} [\arX{1710.07759}].
\bibitem{Mirzabekian:1995ck} A.G.\ Mirzabekian and G.A.\ Vilkovisky, \tia{The one loop form-factors in the effective action, and production of coherent gravitons from the vacuum} \doinn{10.1088/0264-9381/12/9/006}{Classical Quantum Gravity}{12}{2173}{1995} [\oarX{hep-th/9504028}].
\bibitem{BMMS} F.\ Briscese, A.\ Marcianò, L.\ Modesto, and E.N.\ Saridakis \tia{Inflation in (super-)renormalizable gravity} \doin{10.1103/PhysRevD.87.083507}{Phys.\ Rev.}{D}{87}{083507}{2013} [\arX{1212.3611}].
\bibitem{BMT}  F.\ Briscese, L.\ Modesto, and S.\ Tsujikawa, \tia{Super-renormalizable or finite completion of the Starobinsky theory} \doin{10.1103/PhysRevD.89.024029}{Phys.\ Rev.}{D}{89}{024029}{2014} [\arX{1308.1413}].
\bibitem{Modesto:2013ioa} L.~Modesto, \tia{Super-renormalizable gravity}, \procsinm{\href{http://dx.doi.org/10.1142/9789814623995_0098}{\bf The Thirteenth Marcel Grossmann Meeting}}{K.\ Rosquist and R.T.\ Jantzen}{World Scientific}{Singapore}{2015} [\arX{1302.6348}].
\bibitem{Dona:2015tra} P.\ Don\`a, S.\ Giaccari, L.\ Modesto, L.\ Rachwa\l, and Y.\ Zhu, \tia{Scattering amplitudes in super-renormalizable gravity} \doij{10.1007/JHEP08(2015)038}{J.\ High Energy Phys.}{1508}{038}{2015} [\arX{1506.04589}].
\bibitem{Koshelev:2017ebj} A.S.\ Koshelev, K.S.\ Kumar, L.\ Modesto, and L.\ Rachwa\l, \tia{Finite quantum gravity in dS and AdS spacetimes} \doin{10.1103/PhysRevD.98.046007}{Phys.\ Rev.}{D}{98}{046007}{2018} [\arX{1710.07759}].
\bibitem{Mag07} M.\ Maggiore, \book{Gravitational Waves, Vols.\ 1 and 2}{Oxford University Press}{Oxford}{UK}{2007}.
\bibitem{MFB}   V.F.\ Mukhanov, H.A.\ Feldman, and R.H.\ Brandenberger, \tia{Theory of cosmological perturbations} \doinn{10.1016/0370-1573(92)90044-Z}{Phys.\ Rept.}{215}{203}{1992}.
\bibitem{KS3}   V.A.\ Kosteleck\'y and S.\ Samuel, \tia{Collective physics in the closed bosonic string} \doin{10.1103/PhysRevD.42.1289}{Phys.\ Rev.}{D}{42}{1289}{1990}.
\bibitem{DHHJ}  N.\ Dalal, D.E.\ Holz, S.A.\ Hughes, and B.\ Jain, \tia{Short GRB and binary black hole standard sirens as a probe of dark energy} \doin{10.1103/PhysRevD.74.063006}{Phys.\ Rev.}{D}{74}{063006}{2006} [\oarX{astro-ph/0601275}].
\bibitem{NHHDS} S.\ Nissanke, D.E.\ Holz, S.A.\ Hughes, N.\ Dalal, and J.L.\ Sievers, \tia{Exploring short gamma-ray bursts as gravitational-wave standard sirens} \doinn{10.1088/0004-637X/725/1/496}{Astrophys.\ J.}{725}{496}{2010} [\arX{0904.1017}].
\bibitem{CaNi}  S.\ Camera and A.\ Nishizawa, \tia{Beyond concordance cosmology with magnification of gravitational-wave standard sirens} \doinn{10.1103/PhysRevLett.110.151103}{Phys.\ Rev.\ Lett.}{110}{151103}{2013} [\arX{1303.5446}].
\bibitem{Tam16} N.\ Tamanini, C.\ Caprini, E.\ Barausse, A.\ Sesana, A.\ Klein, and A.\ Petiteau, \tia{Science with the space-based interferometer eLISA. III: Probing the expansion of the universe using gravitational wave standard sirens} \doij{10.1088/1475-7516/2016/04/002}{JCAP}{1604}{002}{2016} [\arX{1601.07112}].
\bibitem{Bizon:2011gg} P.~Bizo\'n and A.~Rostworowski, \tia{Weakly turbulent instability of anti-de Sitter space} \doinn{10.1103/PhysRevLett.107.031102}{Phys.\ Rev.\ Lett.}{107}{031102}{2011} [\arX{1104.3702}].
\bibitem{Craps:2014vaa} B.~Craps, O.~Evnin, and J.~Vanhoof, \tia{Renormalization group, secular term resummation and AdS (in)stability} \doij{10.1007/JHEP10(2014)048}{JHEP}{1410}{048}{2014} [\arX{1407.6273}].
\bibitem{Craps:2015jma} B.~Craps and O.~Evnin, \tia{AdS (in)stability: an analytic approach} \doinn{10.1002/prop.201500067}{Fortsch.\ Phys.}{64}{336}{2016} [\arX{1510.07836}].
\bibitem{Bizon:2015pfa} P.~Bizo\'n, M.~Maliborski, and A.~Rostworowski, \tia{Resonant dynamics and the instability of anti--de Sitter spacetime} \doinn{10.1103/PhysRevLett.115.081103}{Phys.\ Rev.\ Lett.}{115}{081103}{2015} [\arX{1506.03519}].
\bibitem{BMS}   T.\ Biswas, A.\ Mazumdar, and W.\ Siegel, \tia{Bouncing universes in string-inspired gravity} \doij{10.1088/1475-7516/2006/03/009}{J.\ Cosmol.\ Astropart.\ Phys.}{0603}{009}{2006} [\oarX{hep-th/0508194}].
\bibitem{BKM1}  T.\ Biswas, T.\ Koivisto, and A.\ Mazumdar, \tia{Towards a resolution of the cosmological singularity in non-local higher derivative theories of gravity} \doij{10.1088/1475-7516/2010/11/008}{JCAP}{1011}{008}{2010} [\arX{1005.0590}].
\bibitem{CMNi}  G.\ Calcagni, L.\ Modesto, and P.\ Nicolini, \tia{Super-accelerating bouncing cosmology in asymptotically-free non-local gravity} \doin{10.1140/epjc/s10052-014-2999-8}{Eur.\ Phys.\ J.}{C}{74}{2999}{2014} [\arX{1306.5332}].
\end{thebibliography}
\end{document}